\documentstyle [12pt]{article}
\def\hang{\par\noindent\hangindent 20pt}
\def\double{\baselineskip=24pt}

\def\ref{\par\noindent\hangindent 20pt}
\begin{document}

\vskip 1.5cm

\titlepage

\begin{center}
{\bf SUBSTRUCTURE IN CLUSTERS AND CENTRAL GALAXY PECULIAR VELOCITIES}
\medskip
\vskip 1.2cm
Christina M. Bird \\
\medskip
{\it Department of Physics and Astronomy \\
University of Kansas \\
Lawrence, KS 66045-2151} \\
\medskip
Internet:  tbird@kula.phsx.ukans.edu\\
\end{center}

\medskip

\begin{abstract}

Formation theories for central dominant galaxies in clusters require them
to be located at the minimum of the cluster gravitational potential.
However, 32\% (8 out of 25)
of the clusters with more than 50 measured redshifts have
central galaxies with significant velocity offsets (with respect to other
cluster members).  By studying their velocity distributions
and correlations between velocity and position, I show that the presence
of a large peculiar velocity is strongly correlated with the presence of
substructure in these massive systems.  About 85\%
(21 of 25) of all well-studied
clusters show some evidence for substructure, in contrast to
the 30-40\% found when using only galaxy or gas distributions.

The correlation between substructure and central galaxy location
verifies the hypothesis of Merritt (1985) and Tremaine (1990) that
high peculiar velocities are indicative of recent merger events between
less-massive systems of galaxies.  Dynamical friction should act quickly
to pull the central galaxy, the most massive discrete object in a cluster,
to the minimum of the potential.  The less-massive galaxies retain
information about their primordial subclusters for a longer period of time.

I use an objective partitioning algorithm to assign cluster galaxies to
their host subclumps.  When galaxies are allocated in this fashion to their
subclusters, 75\% of the significant velocity offsets are eliminated.
Only 2 out of the 25 clusters have central galaxies which are not
centrally-located when substructure is considered in the analysis.

\end{abstract}

\medskip

\noindent{Accepted for publication in the {\it Astronomical Journal}

\bigskip
\noindent
{\it Subject headings:}  galaxies:  clusters ---
galaxies:  elliptical and lenticular, cD --- galaxies:  clustering

\vfill

\newpage

\double

\begin{center}
{\bf 1. INTRODUCTION}
\end{center}

As Edwin Hubble pointed out in {\it The Realm of the Nebulae}, galaxies
come in many shapes and sizes, or more accurately in many morphologies
and masses.  Ellipticals, spirals and irregulars can be found in every
extragalactic environment, from low-density or field regions on up to
densely-populated rich clusters.
Only one type of galaxy
seems to {\it require} the high-density environment.  Giant elliptical
galaxies are never found in the field.  Because they appear to be
preferentially located in the density maxima of clusters of galaxies
(Beers \& Geller 1983), they are often
referred to as {\it central} or {\it central dominant} galaxies.

The original definition of a central dominant galaxy (Mathews, Morgan \&
Schmidt 1964) require the galaxy to be surrounded by an extended luminous
envelope.
Many authors have
extended this definition to include any giant elliptical which is centrally
located (cf. Morgan, Kayser \& White 1975; Albert, White \& Morgan 1977).
Malumuth \& Kirshner (1985) found that central galaxies have a shallower
surface brightness profile than ellipticals which are not brightest cluster
members.
Tremaine (1990) defines a central galaxy as a galaxy located at the minimum
of the cluster potential well, which is itself identified using galaxy or
gas density maxima.  This is the definition I will adopt in this work.

Both
optical and X-ray studies of luminosity functions suggest that central
galaxies
are not an extension of the ``normal'' cluster population of ellipticals
(Malumuth \& Kirshner 1985; Morbey \& Morris 1983;
Edge 1991), although they appear to lie close to the same
fundamental plane
which relates surface brightness, velocity dispersion, and radius
(Oegerle \& Hoessel 1991).
Measurements of the stellar velocity dispersions of central galaxies
(and in some cases, their X-ray gas) indicate that
they are the most massive discrete objects within clusters.  For instance,
Mould et al.\ (1990) use observations of the globular cluster system around
M87, the centrally-located giant elliptical in the Virgo Cluster and find
that its mass is $\sim 10^{13}~{\rm M}_{\odot}$, much greater than that of
typical ellipticals.

The origin of spiral and elliptical morphologies in ``normal'' cluster
galaxies remains a mystery.  The extreme properties of central galaxies
suggest that their formation mechanisms must also be unusual, and probably
involve their priveleged location within the cluster.  Three
mechanisms are generally considered likely for the growth of cDs within
the cluster environment:

\smallskip
\hang{$\bullet$~~
In a relaxed cluster with a sufficiently high central gas density,
intracluster gas will gradually cool and condense
into the bottom of the potential well (Fabian, Nulsen \&
Canizares 1984). This infall may trigger
star formation in the central galaxy.  Optical photometry and
H-$\alpha$ imaging (McNamara \& O'Connell 1992) suggest that this is an
unlikely formation mechanism for the nucleus of the galaxy, but the effect
of the gas infall on the evolution of a centrally-located galaxy is
likely to be important.}

\smallskip
\hang{$\bullet$~~Cluster member galaxies which pass through the center of the
cluster may
be tidally stripped, thereby providing material to be accreted onto
a cD. This may explain growth of the envelope, but cannot explain the
central galaxy itself, since the velocity dispersion of cluster members
is nearly three times that of stars in the cD.}

\smallskip
\hang{$\bullet$~~
The cD may form as the product of mergers, capture (as opposed to
stripping) of less massive galaxies into the bottom of the potential well.}

\noindent
Distinguishing between these formation mechanisms probably requires both
detailed models of the physical processes involved and careful distinction
between classical ``cD-type'' galaxies (which are centrally-located and
possess faint extended haloes), D galaxies (which are centrally-located
but have normal surface brightness profiles) and giant ellipticals (which
are not centrally located and have normal surface brightness profiles).

These formation models
have been extensively studied both analytically and numerically
(Merritt 1984, 1985; Richstone \& Malumuth 1983; Malumuth \& Richstone 1984;
Malumuth 1992), under the important assumption that the central galaxies
forms {\it after} the cluster virializes.  Although evaluating them requires
photometric and morphological distinctions which are beyond the scope of this
dynamical study, they do share one important feature.
Each of these scenarios requires that the ``proto-cD'' be located at
the minimum of the cluster gravitational potential for the formation process to
occur sufficiently rapidly, which in turn
implies that any peculiar velocity of the central galaxy relative to
the cluster should be small.  Despite this,
several well-studied clusters have been
cited as having significant cD velocity offsets (e.g. A2670, Sharples, Ellis
\& Gray 1988; A1795, Hill et al.\ 1988;
A85 and A2052, Malumuth et al.\ 1992; A2107, Oegerle \& Hill 1992).
This poses a serious difficulty for
the formation models.  In particular, Malumuth (1992) showed that the
formation of cD galaxies in mergers after cluster virialization could not
reproduce the observed distribution of peculiar velocities in clusters.

Merritt (1985) and Tremaine (1990) suggest that, rather than forming after a
rich system has collapsed, cD galaxies probably form before or during the
virialization of rich clusters.  Then the host clusters themselves merge and
form an ``evolutionary sequence'' of morphologies:  clusters with two
dominant galaxies (Coma/A1656); systems with dumbbell galaxies (A400, A3266,
A3391); and finally, rich relaxed clusters (A2029).  Optical and X-ray
observations (Beers et al.\ 1991; Edge et al.\ 1992) also provide evidence
for these hierarchical models of cluster formation.  The hypothesis
qualitatively explains many of the features of cD galaxies (see Tremaine 1990
for a summary), including the existence of a handful of systems with high
peculiar velocities.  Dynamical friction acts to pull a cD galaxy into the
center of the gravitational potential of two merging clusters on a relatively
short time-scale, compared to other galaxies in the system, because the mass
of the central galaxy is so high.  The remaining galaxies in the
infalling cluster are much less massive, so their distribution in both position
and velocity space should reflect their origin in a distinct subcluster even
after the cD is at rest in the ``new'' potential. This may lead to a high
value for the measured peculiar velocity (this was also pointed out by Hill
et al.\ 1988), either because the substructure leads to an incorrect estimate
of the velocity location, creating an apparent velocity offset, or because
the central galaxy has not yet had time to settle into the potential
minimum.  The recent merger is detectable as
substructure in the galaxy position and velocity distributions.  For instance,
{\it ROSAT} images of Coma suggest that the gE galaxy
NGC 4839 in
the southwestern clump is displaced from its subcluster's gas, which may be
due to the galaxy's interaction with the massive central system
(White et al.\ 1993).

The development of robust hypothesis tests for substructure in clusters
(Dressler \& Shectman 1988; Bird \& Beers 1993; Bird 1993), when combined
with the availibility of large cluster datasets in the literature,
provides us with the first opportunity to test this hierarchical model
of cluster formation.  These tests are designed to quantify
any structure in the velocity or galaxy distributions as well as correlations
between velocity and position.  By analyzing a cluster dataset using all
available information, these diagnostics maximize the information we can
obtain from a relatively small number of datapoints.

Deviations from Gaussian in a cluster's velocity distribution may imply
anisotropic galaxy orbits and/or mixing of two or more distinct subpopulations
of galaxies.  They can be detected using a variety of
techniques:  those based on moments of the distribution (Fitchett \& Merritt
1988; Bird \& Beers 1993), those based on gaps (Bird \& Beers 1993) or those
determined from a Gauss-Hermite expansion (Zabludoff et al.\ 1993).
Computer simulations suggest that the classical coefficients of skewness
and kurtosis are preferable as diagnostics of small deviations from Gaussian
than the other techniques (Bird \& Beers 1993), and so they will be
applied herein.

Well-formulated, powerful hypothesis tests for substructure in the galaxy
distribution on the sky
are more difficult to define than the 1-D diagnostics.  The
two-dimensional tests of mirror symmetry (the $\beta$-test), angular
separation and density
contrast (West et al.\ 1988) are shown by Bird (1993) to be
inherently insensitive to the structures they claim to detect.  Therefore
the only 2-D test applied to the cluster dataset is the Lee statistic,
introduced to the astronomical literature by Fitchett (1988).  The Lee
statistic quantifies the probability that two equal-sized groups provide
a better description of the galaxy distribution than does a one-group fit.

Computer simulations of cluster formation and theoretical prejudice
predict that a dynamically evolved cluster of galaxies will possess a
roughly isothermal halo.  This is the basis of the nearest neighbor
tests for substructure, which are sensitive to local correlations between
velocity and position.  Although the centroid shift test (or
$\alpha$-statistic, West \& Bothun 1990) is less sensitive than
the cumulative deviation or $\Delta$-test of Dressler \& Shectman (1988b;
also Bird 1993b), for consistency with other cluster analyses I apply both
of them to the cluster dataset.  In addition, I introduce the
$\epsilon$-statistic in this context.

Of the 40 or so galaxy clusters with
more than 50 measured redshifts (the minimum number required for a
robust statistical analysis; Bird 1993), 25 have brightest cluster members
of morphological type `gE,' `D' or `cD.'  These systems are presented
in Table 1.  To be included in the cluster dataset, a galaxy must have a
velocity which is within 3 S$_{BI}$ (defined below) of the average
velocity of the system
(this criterion defines the ``naive'' velocity filter given in column 3) and
must lie within 1.5 $h^{-1}$ Mpc of the central galaxy.  (Here I have taken
$H_0 = 100 h$ km s$^{-1}$ Mpc $^{-1}$.)
The following quantities are listed in Table
1:  column (1), the cluster name; column (2), the cluster right ascension
and declination, found from the position of the central galaxy (see Beers
\& Tonry 1986); (3) the velocity filter and number of galaxies meeting
the membership criteria; (4) the recessional velocity, estimated with
the biweight estimator of location C$_{BI}$ (Beers, Flynn
\& Gebhardt 1990), and the 90\%
confidence intervals about that value; (5) the velocity dispersion,
determined using the biweight estimator of scale S$_{BI}$ (Beers, Flynn \&
Gebhardt 1990),
and its associated 90\% confidence intervals; (6) the line-of-sight
velocity of the central galaxy and the error in that quantity, taken from
the original source; (7) the source code.  Positions and measurement
uncertainties are taken from the original sources given in column 7.

In Section 2, I review the robust method of assessing a central galaxy's
location within a cluster (Gebhardt \& Beers 1991) and apply this
technique to the cluster dataset.  Hypothesis tests for substructure in
clusters are presented in Section 3, and are used to study the clusters'
velocity and galaxy distributions.  In Section 4, I describe the objective
partitioning algorithm and apply it to the cluster dataset.
This work is summarized, conclusions are presented
and future work described in Section 5.

\begin{center}
{\bf 2.  PECULIAR VELOCITIES OF CENTRAL GALAXIES}
\end{center}

Quintana \& Lawrie (1982) presented the first systematic study of
peculiar velocities and concluded that central galaxies were preferentially
(although not exclusively)
located at the center of their host cluster's gravitational potential.
Other recent
studies include Sharples, Ellis \& Gray (1988), Zabludoff et al.\ (1990) and
Malumuth et al.\ (1992).
Each of these studies used slightly different
definitions of the central galaxy offset velocity and evaluated significance
in a non-robust manner, and concluded that a large fraction of clusters
had central galaxies with significant peculiar velocities.
But the results for different clusters are hard to compare to each other
directly, because
of the varying definitions.  Gebhardt \& Beers (1991) point out that many
claims of ``speeding cDs'' disappear when a rigorous statistical technique
is applied to cluster datasets.  In this section, I will review their
physically-meaningful and statistically-robust measure of a central
galaxy's peculiar velocity with respect to its host cluster.  Intuitively,
the physical idea is simple.
Is the velocity of a cluster's
central galaxy significantly displaced from the central location of the
velocity distribution as a whole?  In operation, the definition is a little
more subtle.

In their studies of A1795 and A2670, Hill et al.\ (1988) and Sharples, Ellis
\& Gray (1988)
quantify the significance of a central galaxy's
velocity by scaling it to measurement uncertainties
in the location of the cluster velocity distribution, here
called ${\sigma}_m$, and the uncertainty in the velocity
of the central galaxy, ${\sigma}_l$:
\begin{equation}
Z_{T} = \frac{ \delta v }{ ({\sigma}_l^2 + {\sigma}_m^2)^{1/2} }
\label{eq:ZT}
\end{equation}
This definition is also adopted by Oegerle \& Hill (1992) and Hill \&
Oegerle (1993) in their systematic studies of rich clusters, although
these authors used the biweight estimator of scale $S_{BI}$ rather than
the standard deviation $\sigma$ to estimate the uncertainty in the cluster
mean velocity.  Unfortunately, for non-Gaussian distributions,
there is no simple analytical relationship between {\it any} estimator of the
scale of the distribution and the uncertainty in its velocity location
(cf.\ Beers, Flynn \& Gebhardt 1990).  This may lead to an overestimate of
the significance of a given peculiar velocity.  About 40\% of well-studied
clusters have velocity distributions which are inconsistent with Gaussian
(Bird 1993; this work), implying that use of the analytical relation between
dispersion and uncertainty in the mean is hazardous at best.
Even when the measured velocity
distribution is consistent with Gaussian, the difference between bootstrapped
confidence intervals (which consider sampling errors and uncertainties in
the individual measurements themselves) and confidence intervals calculated
analytically (assuming a Gaussian probability distribution for the sample
distribution) can be as large as a factor of two (Beers, Flynn
\& Gebhardt 1990; see
also Mo, Jing \& Borner 1992 for a discussion of the bootstrapping technique).
Our ability to accurately measure the cluster average velocity
should be one of the factors in deciding whether or not a central galaxy has
a {\it significant} peculiar velocity (with respect to the rest of the
cluster).

Additionally, from a physical point of view
the issue is not: {\it ``Are the velocities consistent within the errors?''}
but rather: {\it ``Are the velocities consistent within $S_{BI}$?''}
because we are interested in the probability of drawing the cD velocity
from a distribution with a dispersion $S_{BI}$. To be physically meaningful,
the peculiar velocity should be scaled to the cluster velocity dispersion,
$S_{BI}$.

Gebhardt \& Beers (1991) define a robust measure of
the physical significance of a cD peculiar velocity:
\begin{equation}
Z_{GB} = \frac{ v_{CD} - C_{BI} }{ S_{BI} }
\label{eq:ZGB}
\end{equation}
This quantity is hereafter called the ``$Z$-score'' for a cluster.
The confidence intervals about this quantity are determined directly
from the observational data, using a bootstrap resampling routine.
To consider a velocity offset significant, the 90\%
confidence intervals of its $Z$-score should not bracket zero; this convention
is chosen for consistency with the significance levels of the substructure
diagnostics.

In Table 2
are presented the $Z$-scores for the 25 clusters in the database.
Column (1) is the cluster name,
(2) the peculiar
velocity and (3) the $Z$-score. All velocities are given in km s$^{-1}$,
and errors are taken from the original sources (listed in
Table 1).  In this table and hereafter, checkmarks after a cluster's name
indicates that the cluster possesses a statistically-significant peculiar
velocity.

32\% of the cD clusters (8 out of 25)
have significant peculiar velocities. [A85 does not, even when the foreground
group is excluded from the cluster membership,
in contradiction to the conclusions of Malumuth et al.\ (1992).  This is
apparently due to the
more rigorous definition of significance used herein, as first found by
Gebhardt \& Beers (1990) using a smaller dataset.]
This high fraction
of significant velocity offsets is a major problem for all three formation
scenarios.  If this result is an indication that a third of all central
galaxies are significantly displaced from the minimum of their cluster
potential, then not only are formation theories in trouble, but tidal stripping
should have shredded the low surface brightness envelopes in these systems
(Merritt 1984).

Malumuth et al.\ (1992) propose that hierarchical clustering will produce a
correlation between peculiar velocity and cluster richness, such that high
peculiar velocities only occur in high velocity dispersion clusters. They claim
this is because, in hierarchical models, the richest, most massive
 systems (on average)
will have undergone the most recent merger event. This may not yet
have been erased through dynamical evolution, and will be reflected in a large
peculiar velocity.  Additionally, in clusters
which still retain distinct velocity populations, the global velocity
dispersion is often over-estimated, which further reinforces this
correlation.

In Figure 1, I present the Malumuth et al.\ (1992) correlation using the rich
clusters from this study, as well as poor clusters with cD brightest cluster
members.  These are taken from an optical study by Beers et al.\ (1993).
When the sample is extended over the largest statistically well-sampled
range in velocity dispersions possible, the correlation that Malumuth et al.\
find is confirmed.  The four crosses mark low-mass systems (MKW9, $N_{gal}=$
14; MKW10, $N_{gal}=$ 16; MKW3s, $N_{gal}=$ 10; MKW4s, $N_{gal}=$17) from
Beers et al.\ which appear to violate the relation.  Even for the lower
numbers of redshifts measured for the poor clusters, the use of bootstrapping
to determine the confidence intervals about $S_{BI}$ and $Z$ makes the
estimators statistically stable for most systems.  In these cases, however,
it is possible that the cluster is not evenly sampled, and that more data
would reduce their peculiar velocities.  If these points are excluded,
Figure 1 certainly supports the hypothesis that high peculiar velocities
are only found in high velocity dispersion systems.

\medskip

\begin{center}
{\bf 3.  SUBSTRUCTURE DIAGNOSTICS}
\end{center}

In order to quantify the level of substructure present in a given cluster,
we must first define what a cluster with {\it no} structure is.
Theoretical and computational models of cluster formation predict that
virialized systems should possess Maxwellian or near-Maxwellian
velocity distributions (Saslaw et al.\ 1990; Ueda et al.\ 1993).
Under this prediction,
the line-of-sight velocity distributions observed for rich systems of galaxies
are expected to be Gaussian (or perhaps, longer-tailed than Gaussian).
Unfortunately, predictions for the distribution of galaxies or gas in
projection on the sky are not as clear-cut.  It does seem reasonable that
a relaxed system should be single-peaked in number density, and spherically
or elliptically symmetric.  Finally, in a cluster with an
isothermal dark matter halo, no correlation between position and velocity of
the member galaxies is expected.  In other words, if cluster galaxies exist
within a common isothermal envelope of dark matter---the {\it simplest}
assumption
about cluster structure, and one which is consistent with current X-ray
observations and gravitational lensing experiments---the velocity dispersion
is
independent of radius within the cluster.  Computer simulations suggest
that tidal interactions between merging subclumps may
drive the gravitational potential quickly toward isothermality (Roettinger
Burns \& Loken 1993; Katz \& White 1993).

Therefore, the substructure diagnostics use the above predictions as
definitions of their null hypotheses.  To study deviations from Gaussian
in the cluster velocity distributions, I have applied the coefficients
of skewness and kurtosis (evaluated using the ROSTAT statistics packages
graciously provided by Timothy Beers).  Skewness is a measure of the
asymmetry of the distribution.  For a dataset in which the central location
and scale are not known {\it a priori}, the skewness is defined as

\begin{equation}
a_3  =  {\sum_{i=1}^N (x_i - C_{\bar x})^3 \over {NS_x^3}}
\label{eq:skew}
\end{equation}
\noindent
Here C$_{\bar x}$ is the average value of the data, and S$_x$ is their
standard deviation.

The kurtosis quantifies the relative population of the tails of the
dataset compared to its central region.  The Gaussian is defined to be
neutrally elongated.  If a distribution
is peakier than Gaussian and/or its tails are heavier than expected, it
exhibits positive kurtosis, or is leptokurtic.  In this case one might
conclude that contamination by outliers, either interlopers in the system
or data with unacceptably high errors, is a problem.
A distribution which is less
peaky than Gaussian and/or has light tails exhibits negative kurtosis, or is
platykurtic.
A boxy distribution may indicate that the system is multimodal, consisting
of overlapping distinct populations.

The kurtosis is the fourth moment of
the distribution about its mean:

\begin{equation}
a_4  =  {\sum_{i=1}^N (x_i - C_{\bar x})^4 \over {NS_x^4}} - 3.
\label{eq:kurt}
\end{equation}
\noindent
The factor of $S_x^4$ in the denominator
makes the estimator scale-invariant.  Evaluating the
significance of the skewness and kurtosis is particularly simple.  It is
equal to the probability that a Gaussian distribution would have a
value equal to that obtained for the dataset under consideration.

Fitchett (1988) introduced to the astronomical literature a maximum
likelihood test of the galaxy distribution, called the Lee statistic
or $L_{rat}$ (Lee 1979).
The Lee statistic tests a cluster (or any 2-D dataset) for
the presence of two equal-sized groups.  Its
great advantage over the 2-D diagnostics discussed in the Introduction
is that it is a well-defined hypothesis test, with a
null hypothesis of one group vs. the alternative of two groups being present
in the data.  Certainly many clusters appear to have bimodal galaxy
distributions (Geller \& Beers 1982), so at first glance this seems to be a
reasonable diagnostic.  The algorithm for calculating $L_{rat}$ is detailed
and too long to be presented here, but may be found in Fitchett (1988)
and Bird (1993).

To determine whether a value of $L_{rat}$ is a rejection of the null
hypothesis for a particular cluster, Monte Carlo
simulations of random distributions are generated from the original datasets
by scrambling the declinations.  500 simulations are performed for each
cluster.  The significance level is the probability that a random
dataset will have as high a value of $L_{rat}$ as the observed cluster has.

The importance of local correlations between velocity and position was
first exploited by Dressler \& Shectman (1988).  Their {\it cumulative
deviation} or $\Delta$-test measures the
deviation from the global average velocity and global velocity dispersion of
each individual galaxy in the cluster and its ten nearest neighbors:
\begin{equation}
\Delta \equiv \sum_{i=1}^{N_{gal}}~\delta_i^2~={11 \over S_{x^2}(gl)^2}
{}~\sum_{i=1}^{N_{gal}}~[(C_{\bar{v}}(i)~-~{C_{\bar v}}(gl))^2~+~(S_{v^2}(i)~-~
S_{v^2}(gl))^2]
\label{eq:delta}
\end{equation}
The subscript $gl$ refers to the kinematic quantity evaluated for the
entire (or $gl$obal) cluster dataset.  The subscript $i$ refers to the
quantity evaluated for the nearest neighbor group.
$\Delta$ is normalized, under the assumption that the velocity dispersion
profile is flat for the (spatial) region under study, by generating Monte
Carlo realizations of the cluster and comparing their $\Delta$-values to that
of the observed dataset.

Similarly, West \& Bothun (1990) define a test which measures the sensitivity
of position centroid measurements to selection of galaxies by velocity range.
The position of
each galaxy in a given cluster dataset is weighted by the
reciprocal of the velocity dispersion
of its 10 nearest neighbors in projection on the sky:
\begin{equation}
w_i = {1 \over \sigma_i}
\end{equation}
Then for each galaxy
$i$, the velocity-weighted position centroid is calculated for itself and its
10 nearest neighbors in {\it velocity space}:
\begin{eqnarray}
x_{ci} = {{\sum^{11}_{i=1} x_i w_i} \over {\sum^{11}_{i=1} w_i}}
\nonumber \\
y_{ci} = {{\sum^{11}_{i=1} y_i w_i} \over {\sum^{11}_{i=1} w_i}}
\label{eq:WB}
\end{eqnarray}
where the $w_i$'s are the weights previously defined.  Finally, the test
statistic is calculated.  The centroid shift is
\begin{equation}
\alpha = {1 \over N_{gal}} \sum^{N_{gal}}_{i=1} [ (x_c - x_{ci})^2 + (y_c -
y_{ci})^2 ] ^ {1 \over 2}
\label{eq:censhift}
\end{equation}
The unweighted $x$- and $y$-centroids for the entire cluster dataset are
designated by $x_c$ and $y_c$ respectively.  The $\alpha$-statistic is
normalized in the same fashion as the $\Delta$-statistic, through repeated
Monte Carlo simulations.

Bird (1993, 1994) points out that using a constant value for the number
of nearest neighbors reduces the sensitivity of these tests to significant
structure.  Silverman (1986) suggests that a value of
\begin{equation}
N_{kern} = N_{gal}^{1 \over 2}
\end{equation}
maximizes sensitivity of the test to significant structure while reducing
sensitivity to fluctuations which are within the Poisson noise.  The
appropriate value of $N_{kern}$ for each galaxy cluster is listed in Table 3.

In both of these substructure diagnostics, the spatial and velocity dimensions
of the cluster data are treated independently:  the $\Delta$-test uses
positions
only to define nearest neighbor groups (hereafter NNGs), the $\alpha$-test
uses the velocities only to select and weight galaxies for the position
centroid calculation.
But position and velocity data both retain information about the degree of
dynamical processing in clusters.  In order to exploit this fact, I define
the $\epsilon$-statistic as
\begin{equation}
\epsilon~=~{1 \over N_{gal}}~\sum_{i=1}^{N_{gal}}~ M_{PME}
\label{eq:eps}
\end{equation}
where $M_{PME}$ is the projected mass estimator (Bahcall \& Tremaine 1981;
Heisler, Tremaine \& Bahcall 1985) calculated for each of the NNGs.
\begin{equation}
M_{PME}~=~\xi~{\left({24 \over \pi GN_{kern}}\right)}~\sum_{j=1}^{N_{kern}}~
v_{zj}^2 r_j
\label{eq:projected}
\end{equation}
The constant $\xi$ depends on the configuration of test particle orbits.
It is here taken to be $4 \over 3$, the value for isotropic orbits.   $v_{zj}$
is the peculiar velocity of a galaxy with
respect to the NNG average velocity, $C_{BI}(i)$, and $r_j$ is its projected
distance from
the NNG center.  The galaxy $i$ for which the NNG is selected does not
contribute
to $M_{PME}$, because it is by definition located at the center of the NNG
and its projected separation is therefore zero.

The $\epsilon$-statistic is normalized in the same way as the $\Delta$-test,
by using the original
cluster dataset.  Random realizations of a cluster with an identical velocity
histogram are obtained by shuffling the velocities and positions of the
observed cluster galaxies.  This Monte Carlo technique effectively removes
any structure in the original data.  It provides an empirical estimate of the
average mass per NNG including Poisson fluctuations in a cluster with a flat
velocity dispersion profile.

A detailed comparison of the three nearest neighbor diagnostics, using
computer simulations to study the power and sensitivity of the tests,
is presented in Bird (1993, 1994).  Study of Table 3 reveals that the 3
diagnostics do not consistently return the same answers for a given dataset.
For instance, A2063 has a marginally significant value of $\Delta$ but a
non-significant $\epsilon$ result.  Computer simulations show that although
the $\epsilon$-statistic is more sensitive to plane-of-the-sky structures
than $\Delta$, the combination of velocity and position data introduces a
larger source of scatter in the Poisson values generated during its
normalization.  This decreases its sensitivity.  In other words,
because the Poisson
fluctuations for the $\epsilon$-test are larger than for the $\Delta$-test,
it is not as sensitive a diagnostic.  Nonetheless, it does sometimes provide
evidence for structure that the other diagnostics miss.  The centroid shift
test suffers from even larger Poisson fluctuations than does $\epsilon$.
Bird (1994) shows that its convoluted algorithm of weights and filters makes
it the least sensitive of the nearest neighbor tests.

In Table 3, I have summarized the results of the substructure
diagnostics for the clusters with high central galaxy peculiar velocity.
I require a significance level of less than 10\% for a rejection of the
null hypothesis with any of these statistics.  These values are given in
the columns to the right of the cluster name.

In contradiction to the commonly-held belief that clusters with central
galaxies have smooth and relaxed morphologies,
these systems have an extremely high frequency of
substructure, as Table 4 demonstrates.
``Low-$Z$'' systems do not have a significant velocity offset; ``high-$Z$''
systems do.  The table summarizes the frequency of substructure detections
for each statistic (0.000 means that no clusters have that sort of structure;
1.000 means they all do).
I have excluded the Lee statistic in the following discussion
because it has no sensitivity to velocity structure.  Since we are trying to
correlate substructure in clusters with a velocity-related property (their
$Z$-score), it makes sense at least at first to consider only diagnostics
which are themselves dependent on (or related to) velocity.  The Lee
statistic, at least as applied herein, is sensitive only to structure on the
sky.  A merger occurring in the plane of the sky is probably taking place
between two subclusters at similar recessional velocities.  In such
systems, no matter how recent the merger the central galaxy will always
appear to be centrally-located, i.e.\ will have a low $Z$-score.  Table 3
shows that in fact low-$Z$ systems do frequently have a positive Lee
statistic.

The high-$Z$ clusters consistently demonstrate a higher frequency of
substructure than do their low-$Z$ counterparts.  In particular, all
of the NNG diagnostics reveal that clusters with a high peculiar
velocity are significantly more likely to reject the null hypothesis of a
relaxed system than those with no large velocity offset.  Also important
is the result that
the only clusters with no indications of substructure (A1809, A2124, and
A2199) have low velocity offsets.
These trends all
lend support to the hypothesis of Merritt (1984, 1985),
Hill et al.\ 1988, Tremaine (1990)
and Malumuth (1992) that central galaxies form in host groups which {\it then}
collapse and virialize to become rich, relaxed clusters.  Although
this idea has been in the literature
for nearly a decade, it has required the compilation of a large dataset of
well-studied clusters and the development of rigorous statistical techniques
presented in this work
to demonstrate that the predicted trend exists in the data.

In the past, 30-40\% of all clusters have been found to demonstrate
substructure
(Geller \& Beers 1982; Jones \& Forman 1984), although Baier (1983)
claimed a much
higher fraction.  These surveys have been
based on tracer particle distributions (of galaxies and gas, respectively)
and have therefore been insensitive to substructure existing in velocity
distributions, that is, along our line of sight.  The frequencies of
rejection of the Gaussian hypothesis for velocity distributions (also
between 30-40\%, see Table 4) imply that the fraction of unrelaxed
clusters may be more than twice as high, $\sim$ 85\%.
Taken another way, only $\sim$15\%
(3 out of 25) of the well-studied clusters in this database show no
indication of substructure in either their velocity or position distributions.
If the cluster sample is extended to include clusters such as A2151 which
do not have a central dominant galaxy (Bird 1993), the fraction of
clusters with no detectable substructure drops to $\sim$8\%.
This substructure may not be dynamically important in every cluster,
but it is certainly sufficient to lend
an added degree of uncertainty to any kinematic or dynamical analyses of
clusters.

\medskip

\begin{center}
{\bf 4.  CORRECTING FOR SUBSTRUCTURE}
\end{center}

Many authors working on the problem of ``speeding central galaxies'' have
proposed substructure as a solution (i.e., Sharples, Ellis \& Gray 1988;
Gebhardt \& Beers 1991; Malumuth et al.\ 1992).  In particular, identifying
the most appropriate membership for the determination of peculiar velocities
was the primary goal of the Malumuth et al.\ (1992) subjective membership
criteria.  They identify cluster members on the basis of
their position within the cluster wedge diagrams and velocity histograms,
also making use of cluster diagnostics such as those described in Section 3.
They admit that this method does not provide a quantitative estimate of
correctness or confidence in their assignments, but they claim that these
criteria are more likely to provide the ``correct'' cluster membership than
the blind application of a 3$\sigma$ clip (Yahil \& Vidal 1977) based on
velocities alone.

Having clearly established the relationship between large offset velocities
and substructure, even in clusters with prominent central galaxies, our
next goal is similarly to identify the
most appropriate {\it host} system for the central
galaxy in the cluster datasets.  To minimize the subjectivity of the analysis,
I have chosen to allocate galaxies to subclusters using an objective
partitioning algorithm called KMM (McLachlan \& Basford 1987; Adams,
McLachlan \& Basford 1993).  KMM is a maximum-likelihood algorithm which
assigns
objects into groups and and assesses the improvement in fitting a
multi-group model over a single group model.  KMM can be applied to data of
any dimensionality.  It has already been applied to one dimensional
astronomical datasets, such as the color/metallicity distributions of
extragalactic globular cluster systems (Ashman \& Zepf 1993;
Zepf \& Ashman 1993; Ostrov et al.\ 1993; Lee \& Geisler 1993).
Ashman, Bird \& Zepf (1994) present an analysis of the statistical behavior
of the algorithm and its use as a tool to detect bimodality in
one-dimensional datasets.  It has been applied to the velocity and position
datasets of galaxy clusters to identify
substructure (A2670, Bird 1994; A548, Davis et al.\ 1994) as the basis of
detailed dynamical analyses.

The KMM algorithm fits a user-specified number of Gaussian distributions to a
dataset and assesses the improvement of that fit over a single Gaussian.
KMM is a hypothesis test of the null hypothesis that the data is consistent
with a unimodal Gaussian probability distribution.  But in addition, it
provides the maximum-likelihood estimate of the unknown $n$-mode Gaussian and
an assignment of objects into groups, unlike the more familiar $\chi^2$ or
KS tests.  This
fitting procedure requires as input a first guess at either the assignments of
the objects into groups or the vector means and covariance matrices of the
model being fit,
as well as the measurements on the objects to be
partitioned.  KMM is most
appropriate in situations where theoretical and/or empirical arguments
indicate that a Gaussian model is reasonable.  This is probably valid in the
case of cluster velocity distributions, where gravitational interactions drive
the system toward Gaussian.  It is less likely to be appropriate for the
projected positions of galaxies on the sky.  But Beers \& Tonry (1986) find
that for any individual cluster of galaxies, the difference in surface
density profile between a Gaussian fit and a de Vaucouleur fit is not large.

The {\it likelihood ratio test statistic}  (hereafter LRTS) provided by KMM
is an estimate of the improvement in going from a 1-mode to a $g$-mode fit.
The significance of any particular LRTS may be estimated by using a $\chi ^2$
distribution, at least in the univariate homoscedastic case, or preferably by
using a
resampling technique such as the bootstrap (McLachlan 1987).  This level is the
fraction of times that as high a LRTS would be calculated if the dataset was
generated under the null hypothesis of a unimodal parent distribution.
That is, the
significance level for any particular KMM run is the chance that the routine
has incorrectly identified a unimodal distribution as $g$-modal.

Selecting the best-fit model is based on the significance level of the
fit,
the algorithm's estimate of the correct
allocation rate (how confident the algorithm is that any individual object
has been assigned to the proper group) and the stability of the means and
dispersions of the Gaussians being fit.
Because the three dimensional diagnostics seem to be the most sensitive
indicators of the presence of substructure, I have chosen to simultaneously
fit the velocity and position data for each of the clusters.  I have fit
two, three and four groups to the datasets.  Few of them contain enough
points to make a fit of more than four groups believable.  In cases where
the LRTS is equal for different models, I have arbitrarily selected the
model with the lowest number of groups to be the best fit.

It is important to emphasize that the problem of estimating the number of
underlying populations in a sample dataset is underspecified.  Therefore
the best-fit models are not mathematically unique.  Subjectivity is
introduced into the use of KMM because the best-fit number of groups is
{\bf not} unique or specified {\it a priori}.
Nonetheless, the algorithm does allow the user
to allocate the data into groups in a statistically consistent way, which
one hopes is more objective than the membership criteria which have been
used in the past.  In addition, use of the KMM algorithm does provide
quantitative measures of membership confidence and goodness of fit, removing
some (although not all) of the subjectivity from the substructure analysis.

The details of the KMM models,
selection of the best-fit partition
and dynamical analyses based on those models
are given in Bird (1993, the high-$Z$ clusters; 1994, A2670).
To save space I have not tabulated the results of each analysis here.
The reader interested in specifically how a best-fit model was chosen for
any particular galaxy cluster is refered to Bird (1993).  In
Table 5, I present for each of the clusters the preferred number
of groups, N$_{gr}$, and the number of galaxies in the group to which the
central galaxy is assigned; the location C$_{BI}$ of the central galaxy's host
subcluster and the 90\% confidence intervals about that value; its dispersion
S$_{BI}$ and the associated confidence intervals; the peculiar velocity
of the central galaxy with respect to its host subsystem; and finally the
recalculated $Z$-score and its 90\% confidence intervals.

Note that for clusters with independent determinations of
substructure (such as Pinkney et al.\ 1993 and Malumuth et al.\ 1992),
KMM has objectively identified the same structures found ``by eye'' and
through use of the Dressler-Shectman statistic.  This provides some reassurance
that despite the uncertainties in the algorithm, in most cases KMM is
allocating galaxies to subclusters in a reasonable way, or at least in a way
which is consistent with previous work.

Examination of Table 5 reveals that
the low-$Z$ systems all retain their low peculiar velocities after
substructure allocation, although in the case of A2199 that may not be
strictly true (Gregory \& Thompson 1984 point out that A2199 is gravitationally
bound to A2197 and therefore, perturbation of the central galaxy from the
center of its host cluster is perhaps unsurprising).
Of the eight high-$Z$ systems, only two (A1736 and A3558) retain their
high offset velocities after substructure is ``corrected'' using the
objective partitioning technique.  To verify that these high offsets are
not an artifact of an unsatisfactory model of the data, I have re-partitioned
the original datasets following each of the KMM models.  In {\it all} of
them, the central galaxy remains significantly displaced from the minimum
of its host system's potential well.  It is possible that the displacement
is physically-meaningful, although in both of these cases the total number
of datapoints is
quite low. More redshifts and positions
will presumably clarify their dynamical situation.

After KMM analysis, only 3 galaxy clusters (A85, A426 and A2199) are found
to consist of only 1 group.  Two of the systems with no detectable
substructure (A1809 and A2124) are classified as bimodal systems, although
in A1809 only 7 galaxies are excluded from the central subcluster.
In all, 9 clusters are bimodal, 7 have three subclusters and 6 have 4.
These numbers are somewhat tentative, given the subjectivity involved in
determining the best-fit model for some cases, but the high frequency of
multimodality is consistent with the high frequency of substructure found
in Section 3.

A careful comparison of the velocity dispersions before and after
substructure allocation reveals that occasionally $S_{BI}$ is larger after
the partition is applied.  This is somewhat counter-intuitive, since in
systems in which the merger is primarily along the line-of-sight,
overlapping velocity distributions can result in an over-estimate of the
dispersion of the system as a whole.  If the KMM best-fit models are
adequate descriptions of the physical structure in the clusters, this
result suggests that partitions based primarily on structure in a velocity
histogram may be misleading, or may not make the most efficient use of the
information available.  On the other hand, it is no doubt true that KMM
has made mistakes in some of its allocations of member galaxies, and that
the ``Occam's razor'' approach to choosing a best-fit model will sometimes
cause mis-assignments.  Nonetheless, in cases where the best-fit models can
be independently verified, either by comparison with X-ray observations or
with other subjective substructure allocations (for instance,
A2634, Pinkney et al.\ 1993; Davis et al.\ 1994), KMM appears to be
providing reliable information.

Having determined the corrected distribution of kinematic and dynamical
properties of galaxy clusters when substructure is included in the analysis,
it is instructive to compare the observations to the previously-cited
theoretical models of cluster formation.  In a forthcoming paper, we will
use these dynamical properties for a revision of the optical and X-ray
correlations.  Here, we can compare the distribution of observed peculiar
velocities to those predicted by Malumuth's (1992) simulations.  First of
all, note that although his simulated clusters each have a velocity
dispersion of 400 km s$^{-1}$, the clusters in his observed distribution
range from $\sim$ 500 km s$^{-1}$ for the poor cluster MKW4 to 1400 km s$^{-1}$
for A2052.  The large scatter in velocity dispersion may be responsible for
the excess of high peculiar velocities when compared to the simulations.
Therefore I normalize the peculiar velocities to the velocity dispersions
before the comparison; that is, I plot the $Z$-score rather than the
peculiar velocity.  This obviously does not change the shape of the
distribution
derived from the simulations but does skew the observed distribution toward
smaller values, making even the uncorrected distribution more consistent
with the theoretical prediction than the original analysis implied.

Unlike the original Malumuth figures, in which large deviations from
the theoretical predictions exist for each bin of the histogram, proper
scaling of the peculiar velocity distribution reduces the inconsistency
even without correction for substructure, as shown in Figure 2.
Using the correctly-allocated
host subclusters, the theoretical and observed predictions are inconsistent
at only a 2$\sigma$ level.  Despite the prevalence of substructure in rich
clusters, perhaps this result implies that central galaxy formation
{\it after} cluster virialization is still a viable formation mechanism.
At least it is not as strongly eliminated from consideration as Malumuth (1992)
maintains.

\medskip
\begin{center}
{\bf 5.  DISCUSSION AND SUMMARY}
\end{center}

The combination of relatively large datasets and robust hypothesis tests
helps to clarify the relationship between central galaxy peculiar velocities
and dynamical evolution in clusters.  Clusters with centrally-located
brightest cluster members have a relatively low frequency of velocity
substructure.  They more often (35\% of the time, 9 of 25) possess quantifiable
structure in their position distributions, which suggests that at least
some of them have formed through ``plane-of-the-sky'' mergers between
less massive systems.  These clusters are presumably at similar
recessional velocities; deconvolving overlapping distributions with the same
means is extremely difficult, and the one-dimensional diagnostics are not
likely to detect this sort of structure.  The $\Delta$- and
$\alpha$-statistics rarely detect structure in low-$Z$ clusters.  The
$\epsilon$-statistic is positive for them about 40\% of the time (7 out of 17).
This is
because its dependence on velocity and position makes it somewhat sensitive to
plane-of-the-sky structure, unlike the other NNG diagnostics (see also Bird
1993b).

High-$Z$ clusters are much likelier to exhibit
structure in their velocity distributions than their low-velocity
counterparts.  They never (or hardly ever) possess positive Lee statistics,
and have a high frequency of detections with the NNG techniques.  As
already seen in the dynamical analysis of A2670 (Bird 1994), these results
are consistent with the merger of subclusters primarily along our line of
sight.  Unfortunately, this orientation makes independent verification
of these subclusters difficult, since X-ray images will not be able
to resolve structure in this orientation.

The substructure analysis confirms the hypothesis of Merritt (1985),
Tremaine (1990) and Malumuth (1992) that high peculiar velocities are an
indication of recent merger events.  The substructure diagnostics are
hypothesis tests of the data's consistency with what we believe a relaxed
system should look like.  Their rejections of the null hypotheses for the
high-$Z$ systems, therefore, support the interpretation that high velocity
offsets are themselves an indication of dynamical turbulence.

It has generally been assumed that highly clumpy, irregular galaxy clusters
have formed recently.  However, some computer simulations suggest
that substructure
may survive for a significant fraction of a Hubble time,
forming quasi-stable structures as the cluster virializes (cf. bimodal
or ``dumb-bell'' structures; Cavaliere \& Colafrancesco 1990).  If this
is true, then the existence of substructure no longer implies that a
cluster has formed recently.  The high mass of central galaxies implies
that dynamical friction will force them to a cluster's potential minimum
on a timescale of a third of a Hubble time or so (Tremaine 1990).  This
is rather quicker than the timescale for less massive galaxies to reach
equilibrium within the potential well.  The correlation between peculiar
velocity and substructure implies that the mergers must have occurred
relatively recently, thereby strengthening the assumption that the
substructure diagnostics are reliable hypothesis tests of a cluster's
dynamical state.

This is not to imply that the substructure tests alone are sufficient to
completely determine the dynamics of a cluster of galaxies.  Small number
statistics even in the largest cluster catalogs introduce an appreciable
source of uncertainty, even when sophisticated statistical techniques are
used.  The importance of reliable X-ray temperature determinations in
merging systems cannot be over-emphasized, as an independent measurement
of the depth of the gravitational potential.

The prevalence of substructure in clusters with central galaxies resolves
the apparent conflict between formation theories for these ultraluminous
galaxies and the high fraction of clusters with displaced central
galaxies.  When substructure is corrected by assigning central galaxies
to their host subclusters, 6 of the 8 significant peculiar velocities
disappear.  This result does not favor any of the
formation theories summarized in the Introduction, although it may
provide some difficulty for the cooling flow model.  Edge et al.\ (1992)
suggest that merger events between subclusters of comparable sizes
disrupt the quasi-equilibrium conditions
necessary for a cooling flow.  These disruptions are relatively short, however,
and may not have a major effect on the amount of mass deposited over a
Hubble time.

Tremaine (1990) proposes that clusters with double-nuclei central galaxies
(the ``dumb-bell galaxies'' defined by Mathews, Morgan \& Schmidt 1964)
may be a profitable subject for this sort of analysis.  In a hierarchical
clustering
sequence, these galaxies are formed during the rapid merger of two
galaxies which are themselves centrally-located within two merging
subclusters.  Because the merging timescale for the two central galaxies is
extremely short, dumb-bell galaxies may be a clear signature of
dynamical youth.  In that case, application of the statistical diagnostics
and partitioning algorithm contained in this work will provide clear
evidence for the interacting subclusters, as we have already seen to be the
case in clusters with classical central galaxies.  This link between dumb-bell
evolution and larger-scale substructure is evident in the Beers et al.\ study
of A400 (1992).  Mergers between central galaxies may also promote the
formation of wide-angle tail radio sources, as seen in A2634 by Pinkney
et al.\ (1993).
It may be possible, for
sufficiently large cluster datasets, to associate the individual dumb-bell
components with their host subclusters.

\medskip

\medskip

\medskip

I am very grateful to Eliot Malumuth and Bill Oegerle for playing devil's
advocate and Timothy Beers for software and a firm grounding in statistical
methodology.  Keith Ashman, Scott Tremaine
and the Canadian Institute for Theoretical
Astrophysics provided support and encouragement during this project.  John
Hill, Ray Sharples, Jason Pinkney and Alan Dressler provided computer
versions of their cluster catalogs, for which I am appreciative.
John Hill improved the text through his careful and prompt referee'ing.
This work was completed in partial fulfillment of the requirements for the
Ph.D program at the University of Minnesota, and supported in part by the
Department of Physics and Astronomy, Michigan State University, and by
NSF grant OSR-9255223 at the University of Kansas.

\newpage

\begin{center}
{\bf REFERENCES}
\end{center}

\hang{Adams, P., McLachlan, G. \& Basford, K.  1993, private communication}
\hang{Albert, C.E., White, R.A. \& Morgan, W.W.  1977, ApJ, 211, 309}
\hang{Ashman, K.M., Bird, C.M. \& Zepf, S.E.  1994, in preparation}
\hang{Ashman, K.M. \& Zepf, S.E.  1993, in Globular Clusters in the Context
of their Parent Galaxies, edited by J.\ Brodie and G.\ Smith (Dordrecht:
Kluwer), p.\ 776}
\hang{Bahcall, J.N. \& Tremaine, S.  1981, ApJ, 244, 805}
\hang{Baier, F.W.  1983, Astron. Nacht., 5, 211}
\hang{Beers, T.C., Flynn, K. \& Gebhardt, K.  1990, AJ, 100, 32}
\hang{Beers, T.C., Forman, W., Huchra, J.P., Jones, C. \& Gebhardt, K.  1991,
AJ, 102, 1581}
\hang{Beers, T.C. \& Geller, M.J.  1983, ApJ, 274, 491}
\hang{Beers, T.C., Huchra, J.P., Kriessler, J. \& Bird, C.M.  1994, in
preparation}
\hang{Beers, T.C., Gebhardt, K., Huchra, J.P., Forman, W. \& Jones, C.
1992, ApJ, 400, 410}
\hang{Beers, T.C. \& Tonry, J.L.  1986, ApJ, 300, 557}
\hang{Bird, C.M.  1993, Ph.D thesis, University of Minnesota and Michigan
State University}
\hang{Bird, C.M.  1994a, AJ, in preparation}
\hang{Bird, C.M.  1994b, ApJ, to appear 20 February}
\hang{Bird, C.M. \& Beers, T.C.  1993, AJ, 105, 1586}
\hang{Cavaliere, A. \& Colafrancesco, S.  1990, in {\it Clusters of Galaxies},
ed. W.R. Oegerle, M. J. Fitchett, and L. Danly, (New York:  Cambridge
University Press), pg. 43}
\hang{Chapman, G., Huchra, J.P. \& Geller, M.  1988, CfA preprint}
\hang{Davis, D.S., Bird, C.M., Mushotzky, R.F. \& Odewahn, S.C. 1994, ApJ,
submitted}
\hang{Dressler, A.  1980, ApJ, 236, 351}
\hang{Dressler, A. \& Shectman, S.  1988a, AJ, 95, 284}
\hang{Dressler, A. \& Shectman, S.  1988b, AJ, 95, 985}
\hang{Edge, A.C.  1991, MNRAS, 250, 103}
\hang{Edge, A.C., Stewart, G.C. \& Fabian, A.C.  1992, MNRAS, 258, 177}
\hang{Fabian, A.C., Nulsen, P.E.J. \& Canizares, C.R.  1984, {\it Nature}, 310,
733}
\hang{Fabricant, D., Kurtz, M., Geller, M., Zabludoff, A., Mack, P. \&
Wegner, G.  1993, AJ, 105, 788}
\hang{Fitchett, M.J.  1988, MNRAS, 230, 169}
\hang{Fitchett, M. \& Merritt, D.  1988, ApJ, 335, 18}
\hang{Gebhardt, K. \& Beers, T.C. 1991, ApJ, 383, 72}
\hang{Geller, M. \& Beers, T.C.  1982, PASP, 94, 421}
\hang{Giovanelli, R. \& Haynes, M.  1985, ApJ, 292, 404}
\hang{Gregory, S.A. \& Thompson, L.A.  1984, ApJ, 286, 422}
\hang{Guzman, R., Lucey, J.R., Carter, D. \& Terlevich, R.J.  1992, MNRAS,
257, 187}
\hang{Heisler, J., Tremaine, S. \& Bahcall, J.N.  1985, ApJ, 298, 8}
\hang{Hill, J.M. \& Oegerle, W.R.  1993, AJ, 106, 831}
\hang{Hill, J.M., Hintzen, P., Oegerle, W.R., Romanishin, W., Lesser, M.P.,
Eisenhamer, J.D. \& Batuski, D.J.  1988, ApJ, 332, L23}
\hang{Hubble, E.  1936, {\it The Realm of the Nebulae}, (New Haven:  Yale
University Press)}
\hang{Jones, C. \& Forman, W.  1984, ApJ, 276, 38}
\hang{Katz, N. \& White, S.D.M.  1993, ApJ, 412, 455}
\hang{Kent, S.M. \& Sargent, W.L.W.  1983, AJ, 88, 697}
\hang{Lee, K.L.  1979, JAmStat, 74, 708}
\hang{Lee, M.G. \& Geisler, D.  1993, AJ, 106, 493}
\hang{Malumuth, E.M.  1992, ApJ, 386, 420}
\hang{Malumuth, E.M. \& Kirshner, R.P.  1985, ApJ, 291, 8}
\hang{Malumuth, E.M. \& Richstone, D.O. 1984, ApJ, 276, 413}
\hang{Malumuth, E.M., Kriss, G.A., Dixon, W.V.D., Ferguson, H.C. \& Ritchie,
C.  1992, AJ, 104, 495}
\hang{Mathews, T.A., Morgan, W.W. \& Schmidt, M.  1964, ApJ, 140, 35}
\hang{McLachlan, G.J.  1987, Appl. Statist., 36, 318}
\hang{McLachlan, G.J. \& Basford, K.E.  1988, {\it Mixture Models}, (New York:
Marcel Dekker)}
\hang{McNamara, R.B. \& O'Connell, R.W.  1992, ApJ, 393, 579}
\hang{Merritt, D. 1984, ApJ, 276, 26}
\hang{Merritt, D. 1985, ApJ, 289, 18}
\hang{Mo, H.J., Jing, Y.P. \& Borner, G.  1992, ApJ, 392, 452}
\hang{Morbey, C. \& Morris, S.  1983, ApJ, 274, 502}
\hang{Morgan, W.W., Kayser, S. \& White, R.A.  1975, ApJ, 199, 545}
\hang{Mould, J.R., Oke, J.B., de Zeeuw, P.T. \& Nemec, J.M.  1990, AJ,
99, 1823}
\hang{Oegerle, W.R. \& Hill, J.M. 1992, AJ, 104, 2078}
\hang{Oegerle, W.R. \& Hoessel, J.G.  1991, ApJ, 375, 15}
\hang{Ostrov, P., Geisler, D. \& Forte, J.C.  1993, AJ, 105, 1762}
\hang{Pinkney, J., Rhee, G., Burns, J.O., Hill, J.M., Oegerle, W.,
Batuski, D. \& Hintzen, P. 1993, ApJ, 416, 36}
\hang{Quintana, J. \& Lawrie, 1982, AJ, 87, 1}
\hang{Richstone, D.O. \& Malumuth, E.M. 1983, ApJ, 268, 30}
\hang{Richter, O.G.  1987, A\&ASuppl, 77, 237}
\hang{Roettinger, K., Burns, J.O. \& Loken, C.  1993, ApJLett, 407, L53}
\hang{Saslaw, W.C., Chitre, S.M., Itoh, M. \& Inagaki, S.  1990, ApJ, 365, 419}
\hang{Sharples, R., Ellis, R. \& Gray, P.  1988, MNRAS, 231, 479}
\hang{Sodre, L., Capelato, H.V., Steiner, J.E., Proust, D. \& Mazure, A. 1992,
MNRAS, 259, 233}
\hang{Teague, P.F., Carter, D. \& Gray, P.M.  1990, ApJSuppl, 72, 715}
\hang{Tremaine, S. 1990, in {\it Dynamics and Interactions of Galaxies}, ed.\
R. Wielen, (Berlin:  Springer-Verlag), pg. 394}
\hang{Ueda, H., Itoh, M. \& Suto, Y.  1993, ApJ, 408, 3}
\hang{West, M.J. \& Bothun, G.D.  1990, ApJ, 350, 36}
\hang{West, M.J., Oemler, A. \& Dekel, A.  1988, ApJ, 327, 1}
\hang{White, S.D.M., Briel, U.G. \& Henry, J.P. 1993, MNRAS, 261, L8}
\hang{Whitmore, B.C. \& Gilmore, D.  1991, ApJ, 367, 64}
\hang{Yahil, A. \& Vidal, N.V.  1977, ApJ, 214, 347}
\hang{Zabludoff, A.I., Huchra, J.P. \& Geller, M.J.  1990, ApJSuppl, 74, 1}
\hang{Zabludoff, A.I., Franx, M. \& Geller, M.J.  1993, CfA preprint}

\newpage

\begin{center}
{\bf FIGURE CAPTIONS}
\end{center}

\hang{Figure 1:  The velocity offset of central dominant galaxies is
plotted against their velocity dispersion $S_{BI}$.  Stars indicate rich
clusters from this study; solid circles indicate poor clusters from Beers
et al\ (1993).  The crosses are the 4 poor clusters with high $Z$-scores
and low velocity dispersions, discussed in the text.  Error bars on $S_{BI}$
are the 90\% confidence intervals given in Table 1.}

\hang{Figure 2:  The observed distributions of peculiar velocities in
rich clusters of galaxies, before and after correction for substructure,
to be compared with Malumuth (1992) Figure 4 and binned in the same
increments.  The upper panels are the
original histograms; the lower panels provide a ``double-root residual''
transformation of the difference between the observed distributions and
Malumuth's simulations (see Gebhardt \& Beers 1991; Ashman, Bird \& Zepf 1994
for a detailed description of the double root residual).  Note that even
without substructure correction, deviations from the predicted distribution
of peculiar velocities are all within 3$\sigma$.  When substructure is
considered in the analysis, only the second bin of the histogram deviates
by 2$\sigma$ from the predicted value.}

\end{document}